\documentstyle{article}
\begin{document}
\vspace{-2cm}
\title{\sc Diffractive Mass Spectra at HERA
	   in the Interacting Gluon Model}
\author{F.O. Dur\~aes$^1$\thanks{e-mail: dunga@uspif.if.usp.br}, \ F.S.
Navarra$^{1}$\thanks{e-mail: navarra@uspif.if.usp.br} \ and \ G.
Wilk$^{1,2}$\thanks{e-mail: wilk@fuw.edu.pl} \\ 
{\it $^1$Instituto de F\'{\i}sica, Universidade de S\~{a}o Paulo}\\
{\it C.P. 66318,  05315-970 S\~{a}o Paulo, SP, Brazil} \\[0.1cm]
{\it$^2$Soltan Institute for Nuclear Studies, 
Nuclear Theory Department}\\
{\it ul. Ho\.za 69, \ 00-681 Warsaw, Poland}}
\maketitle
\vspace{1cm}
\begin{abstract}
We have successfully applied the Interacting Gluon Model (IGM) 
to calculate diffractive mass spectra 
 measured recently in e-p collisions at HERA. We
show that it is possible to treat them in terms of 
gluon-gluon collisions in the same way as was done before for
hadronic collisions. Analysis of available data
is performed. \\

PACS number(s): 13.85.Qk, 11.55.Jy\\

\end{abstract}

\vspace{1cm}
\section{Introduction}

Diffractive scattering processes  are 
related to the large rapidity gap physics usually interpreted
in terms of Pomeron exchange \cite{POMERON}. In hadronic 
diffractive scattering, one of the incoming hadrons emerges 
from the collision only slightly deflected and there is a 
large rapidity gap between it and the other final state particles 
resulting from the other excited hadron. In the standard Regge theory 
diffraction is visualised as due to the Pomeron exchange which 
implies that the excited mass spectrum behaves 
like $1/M^2_X$ and does not depend on the energy \cite{REGGE}. 
The same phenomena can, however, be understood 
without mentioning 
the name of Pomeron by treating $I\!\!P$ as a {\it
preformed colour singlet} object consisting of only a
part of the gluonic content of the diffractive projectile
which is then absorbed by the other hadron \cite{NONPOM}.
One possible realization of this idea was developed recently by us
\cite{IGM97} (for earlier similar attempts see \cite{OTHER}). 

The first test of a theory (or a model) of diffractive dissociation 
(DD) is the ability to properly describe the mass ($M_X$) 
distribution of diffractive systems, which has been measured in many 
hadronic collision experiments \cite{DATA} and parametrized as 
$(M_X^2)^{-\alpha}$ with $\alpha \simeq 1$. 

Very recently diffractive mass spectra have been measured also 
in photoproduction processes at high energies at HERA \cite{HERA}. 
They were interpreted there in terms of standard Regge theory. 
In this work we would like to analyse these data using instead
the Interacting Gluon Model (IGM) which was already successfully used
to describe diffractive mass spectra and their energy dependences in
hadronic reactions \cite{IGM97}. The advantage of such approach is 
the possibility
to check what part of the results is due to the simple implementation
of conservation laws (notice that IGM was designed in such 
a way that the energy-momentum conservation is taken care of before 
all other dynamical aspects - a feature very appropriate for the 
study of all kinds of energy flows). It means that all notions 
to the Pomeron (or $I\!\! P$) in what follows is just symbolic 
representation of the DD process.

\section{IGM picture of a diffractive event}

As mentioned in \cite{HERA}, at the HERA electron-proton collider
the bulk of the cross section corresponds to photoproduction, in
which a beam electron is scattered through a very small angle and
a quasi-real photon interacts with the proton. For such small
virtualities the dominant interaction mechanism takes place via
fluctuation of the photon into a hadronic state which interacts with
the proton via the strong force. High energy photoproduction therefore
exhibits similar characteristics to hadron-hadron interactions.

In Fig. 1 we show schematically the IGM picture of a
diffractive event in a photon-proton collision.
According to it, during the interaction the photon is converted
into a hadronic (mesonic) state and then interacts with the incoming
proton \cite{VMD}. The meson-proton interaction follows then the usual 
IGM picture, namely: the valence quarks fly through essentially 
undisturbed whereas the gluonic clouds of both projectiles interact 
strongly with each other (by gluonic clouds we understand a sort of 
"effective gluons" which include also their fluctuations seen as 
$\bar{q}q$ sea  pairs). The meson looses fraction $x$ of its original 
momentum and gets excited forming
what we call a {\it leading jet} (LJ) carrying $x_L= 1 -x$ fraction
of the initial momentum. The proton, which we shall call here the
diffracted proton, looses only a fraction $y$ of its momentum but
otherwise remains intact \cite{FOOT}.

In the limit $y\rightarrow 1$, the whole available energy is stored in
$M_X$ which then remains at rest, i.e., $Y_X = 0$. For small values
of $y$ we have small masses $M_X$ located at large rapidities $Y_X$.
In order to regard our process as being trully of the DD type we must
assume that all gluons from the target proton participating in the
collision (i.e., those emitted from the lower vertex in Fig. 1) 
have to form a colour singlet. Only then a large rapidity gap will
form separating the diffracted proton (in the lower part of our Fig.
1) and the $M_X$ system (in its upper part), which is the
experimental requirement defining a diffractive event. Otherwise a 
colour string would develop, connecting the diffracted proton and the
diffractive cluster, and would eventually decay, filling the
rapidity gap with produced secondaries. In this way we are
effectively introducing an object resembling closely to what 
is known as Pomeron ($I\!\!P$) and therefore in what follows 
we shall use this notion. The Pomeron may be treated \cite{IS}
 as being composed of partons, i.e., gluons and sea 
$\bar{q}q$ pairs, in much the same way as hadrons, with some 
characteristic distribution functions which have been object
of studies in HERA experiments \cite{zeus,h1}.
 
As usual in the IGM \cite{IGM97} we first start with the function 
$\chi(x,y)$ describing the probability to form a central
gluonic fireball (CF) carrying momentum fractions $x$ and $y$ of the
two colliding projectiles: 
\begin{eqnarray}
\chi(x,y) &=& \frac{\chi_0}{2\pi\sqrt{D_{xy}}}\cdot \nonumber\\
&&\cdot \exp \left\{ - \frac{1}{2D_{xy}}\,\left[
  \langle y^2\rangle (x - \langle x\rangle )^2 +
  \langle x^2\rangle (y - \langle y\rangle )^2 -
  2\langle xy\rangle (x - \langle x\rangle )(y - \langle y\rangle )
  \right] \right\}. \label{eq:CHI}
\end{eqnarray}
For our specific needs in this paper (application to DD events)
where we are mostly interested in the $x$ and $M_X^2$ behaviour 
of the results, it is usefull to present (\ref{eq:CHI}) in the 
form where the $x$-dependence is factorized out:
\begin{equation}
\chi(x,y) = \frac{\chi_0}{2\pi\sqrt{D_{xy}}}\cdot
                \exp \left[ 
		- \frac{\left( y - \langle y\rangle\right)^2}
		       {2\langle y^2\rangle} \right] \cdot
                \exp \left\{ - \frac{\langle y^2\rangle}{2D_{xy}}
                \left[ x - \langle x\rangle - 
		\frac{\langle xy\rangle}
                     {\langle y^2\rangle}
		 (y - \langle y\rangle)\right]^2\right\}.
		 \label{eq:CHINEW}
\end{equation}
In the above equations
\begin{eqnarray}
D_{xy} &=& \langle x^2\rangle \langle y^2\rangle - 
           \langle xy\rangle ^2   
\end{eqnarray}
and
\begin{eqnarray}
\langle x^ny^m\rangle &=& \int_0^1\! dx\,x^n\, \int_0^{y_{max}}\! dy\, 
y^m\, \omega (x,y). \label{eq:defMOM}
\end{eqnarray}
Here $\chi_0$ denotes the normalization factor provided by the
requirement that $ 
 \int_0^1\!dx\, \int_0^1\! dy\, \chi(x,y) \theta(xy - K_{min}^2) = 1
$
with $K_{min} = \frac{m_0}{\sqrt{s}}$ being the minimal inelasticity
defined by the mass $m_0$ of the lightest possible CF and $\sqrt{s}$ is
proton-hadron center of mass energy. In the above
expression $y_{max} = \frac{M^2_X}{s}$. This upper cut-off, not 
present in the non-diffractive formulation of the IGM 
(where $y_{max}=1$), is 
necessary to adapt the standard IGM to DD collisions. It is a
kinematical restriction preventing the gluons coming from the
diffracted proton (and forming our object $I\!\!P$) to carry more
energy than what is released in the diffractive system. 
As it will be seen below, it plays a central 
role in the adaptation 
of the IGM to DD processes being responsable for its proper
$M^2_X$ dependence. The, so called, spectral function 
$\omega(x,y)$ contains all the dynamical inputs of the 
IGM in the general form given by (cf. \cite{DNW93}) 
\begin{equation}
\omega(x,y)\, =\, \frac{\sigma_{gg}(xys)}{\sigma(s)}
   \, G(x)\, G(y)\, \Theta\left(xy - K^2_{min}\right),
   \label{eq:OMEGA}
\end{equation}
where $G$'s denote the effective number of gluons from the
corresponding projectiles (approximated by the respective gluonic 
structure functions) and $\sigma_{gg}$ and $\sigma$ are the gluonic 
and hadronic cross sections, respectively. In order to be more
precise, the function $ G(y) $, as it can be seen in Fig. 1, represents
the momentum distribution of the gluons belonging to the proton subset
called Pomeron and $y$ is the momentum fraction {\it of the proton}
carried by one of these gluons. We shall therefore use the notation
$ G(y) = G_{I\!\!P}(y)$. This function should not be confused with the
momentum distribution of the gluons inside the Pomeron, 
$f_{g/I\!\!P}(\beta)$ (see below).

The moments  $\langle q^n\rangle,~~q=x,y$ (we only require
$n=1,2$) are given by
(\ref{eq:defMOM}) and are the only places where dynamical quantities
like the gluonic and hadronic cross sections appear in the IGM. At this
point we emphasize that we are all the time dealing with a meson
(essentially the $\rho^0$)-proton scattering. However, as was said 
above, we are in fact selecting a special class of events and 
therefore we must choose the correct dynamical inputs in the 
present situation, namely $G_{I\!\!P}(y)$ and the hadronic cross 
section $\sigma$ appearing in $\omega$. 

As pointed out in the introduction, the Pomeron for us is just a
collection of gluons which belong to the diffracted proton. In our 
previous work we have assumed that these gluons behave like all other 
ordinary gluons in the proton and have therefore the same momentum
distribution. The only difference is the momentum sum rule, which for 
the gluons in $I\!\!P$  is
\begin{eqnarray}
\int_0^1\! dy\,y\,G_{I\!\!P}(y)\,=\,p
\end{eqnarray}
where $p\,\simeq\,0.05$ (see  \cite{IGM97} and below )
instead of $ p\simeq 0.5$, which holds for the entire gluon population
in the proton. Alternatively we
may treat the Pomeron structure in more detail and address the question
of its ``hardness'' or ``softness''. In order to make 
contact with the analysis performed by HERA experimental groups we consider 
two possible momentum distributions for the gluons inside  $I\!\!P$ :
\begin{eqnarray}
f^h_{g/I\!\!P}(\beta) &=&  6\,(1\,-\,\beta) 
\end{eqnarray}
and
\begin{eqnarray}
f^s_{g/I\!\!P}(\beta) &=&  6\,\frac{(1\,-\,\beta)^5}{\beta} 
\end{eqnarray}
where $\beta$ is the momentum fraction of the Pomeron carried by the gluons
and the superscripts $h$ and $s$ denote hard and soft respectively. We 
follow here the (standard) notation of ref. \cite{zeus}. We shall use
the Pomeron flux factor given by
\begin{eqnarray}
f_{I\!\!P/p}(x_{I\!\!P}) &=& \frac{1}{x_{I\!\!P}}  
\end{eqnarray} 
where $x_{I\!\!P}$ is the fraction of the proton momentum carried by the
Pomeron and the normalization will be fixed later. Noticing  that
$\beta = \frac{x}{x_{I\!\!P}}$ the distribution $G_{I\!\!P}(y)$ needed 
in eq. (\ref{eq:OMEGA}) is then given by the convolution: 
\begin{eqnarray}
G^{h,s}_{I\!\!P} (y) &=& \int_y^1\! \frac{ dx_{I\!\!P} } {x_{I\!\!P}} \,
f_{I\!\!P/p}(x_{I\!\!P})\,f^{h,s}_{g/I\!\!P}(\frac{y}{x_{I\!\!P}})
\end{eqnarray}
In our calculations we shall also use 
$G_{I\!\!P}(y) = 6 \frac{(1-y)^5}{y}$, the same expression
already used by us before \cite{IGM97}. As it will be seen this choice
corresponds to an intermediate between ``soft'' and ``hard'' Pomeron.
 In the upper leg of Fig. 1 
we assume, for simplicity, the vector meson to be $\rho^{0}$ and take 
$ G^{\rho^{0}}(x) = G^{\pi}(x)$. The fraction of
diffracted nucleon momentum, $p$, allocated specifically to the
$I\!\!P$ gluonic cluster and the hadronic cross section $\sigma$ are
both unknown. However, they always appear in $\omega$ as a ratio
($\frac{p}{\sigma}$) of parameters and different choices
are possible. Just in order to make use of the present
knowledge about the Pomeron, we shall choose
\begin{eqnarray}
\sigma(s) \,&=& \, \sigma^{I\!\!P p} \, = \, a + 
b \, \ln \frac{s}{s_0}
\label{eq:defsig}
\end{eqnarray}
where $s_0 = 1$ $GeV^2$ and $a = 2.6$mb  and $b= 0.01$mb are 
parameters fixed from a previous \cite{IGM97} systematic data
analysis. As it can be seen, $\sigma(s)$ turns out to be a very
slowly varying function of $\sqrt{s}$ assuming values between 2.6
and 3.0 mb, which is a well accepted value for the
Pomeron-proton cross section, and $ p \simeq 0.05 $ 
(cf. \cite{IGM97}). Since the parameter $\frac{p}{\sigma}$ 
has been fixed considering the
proton-proton diffractive dissociation and we are now addressing
the $p\,-\,\rho^{0}$ case we have some freedom to change $\sigma$.
In the following we shall also investigate the effect of small 
changes in the value of $m_{0}$ on our final results. 

Although in the final numerical calculations the above complete 
formulation will be used, it is worthwhile to present 
approximate analytical results in order to illustrate the 
main characteristic features of the IGM. It is straightforward to
show that, keeping only the 
most singular terms in gluon distribution functions, 
i.e., $G(x)= G^{h,s}_{I\!\!P}(x) 
 \simeq \frac{1}{x}$ and only the leading terms in 
$\sqrt{s}$, one finds that for any choice of $\sigma_{gg}$ in
(\ref{eq:OMEGA}):
\begin{itemize}
\item[$(i)$] terms containing the mixed moment $\langle xy\rangle$ 
can be always neglected in comparison to those containing 
$\langle x^2\rangle$ or $\langle y^2\rangle$ (i.e., for example, 
$D_{xy} \simeq \langle x^2\rangle\langle y^2\rangle$); 
\item[$(ii)$] all moments are related in the following way:
\begin{equation}
\langle x^2\rangle \simeq \frac{1}{2}\langle x\rangle ; 
\qquad 
\langle y\rangle \simeq  y_{max}\langle x\rangle ;
\qquad
\langle y^2\rangle \simeq \frac{1}{2}y_{max}\langle y\rangle ;
\label{eq:relmom}
\end{equation}
i.e., all results can be expressed in terms of the
$\langle x\rangle$ moment only;
\item[$(iii)$] the $\langle x\rangle$ moment has the following simple
behaviour depending on the type of $\sigma_{gg}$ chosen in
(\ref{eq:OMEGA}):\\
\begin{equation}
\langle x\rangle \simeq const, \quad 
                 \simeq  \ln\left(\frac{sy_{max}}{m_0^2}\right), \quad
    \simeq \frac{1}{2} \ln^2\left(\frac{sy_{max}}{m_0^2}\right)\qquad
{\rm for}\qquad \sigma_{gg}\simeq \frac{m_0^2}{xys},\quad
                      \simeq const,\quad
    \simeq 
    \ln\left(\frac{xys}{m_0^2}\right),\label{eq:xmom}
\end{equation}
respectively.
\end{itemize}
This allows us to write eq.(\ref{eq:CHINEW}) in a very simple form:
\begin{equation}
\chi(x,y) \simeq \frac{\chi_0}{\pi\, y_{max}\langle x\rangle}\cdot
            \exp \left[ - \frac{(y - y_{max}\langle x\rangle)^2}
                               {y^2_{max}\langle x\rangle} \right]\cdot
            \exp \left[ - \frac{(x - \langle x\rangle)^2}
                               {\langle x\rangle} \right] .
                               \label{eq:CHIAPP}
\end{equation}

As already mentioned, the $\frac{1}{y_{max}}$ term present in 
(\ref{eq:CHIAPP}) can be traced back to the upper cut-off 
$y=y_{max}$ in (\ref{eq:defMOM}) above. Because it is defined 
by the produced diffractive mass, $y_{max} = \frac{M^2_X}{s}$,
it provides then automatically the  $\frac{1}{M^2_X}$ behaviour. 
The other two factors have a much weaker dependence on 
$M^2_X$ and they tend to compensate each other  (they provide,
however, all possible non-trivial energy dependence of DD, cf. 
\cite{IGM97}).

\section{Comparison with experimental data}

The IGM diffractive mass spectrum is given by \cite{IGM97}:
\begin{eqnarray}
\frac{dN}{dM_X^2}\, &=&\, \int_0^1\! dx\, \int_0^1\! dy\, 
                       \chi(x,y)\, \delta\left( M^2_X - sy\right)\, 
                       \Theta \left( xy - K^2_{min}\right)
                       \nonumber \\
                    &=& 
          \frac{1}{s}\, \int_{_{\frac{m_0^2 }{M_X^2}}}^1\! dx\,
          \chi\left( x,y=\frac{M_X^2}{s}\right), \label{eq:RES}
\end{eqnarray} 
or, in the approximate form,
\begin{equation}
\frac{dN}{dM_X^2} \simeq \frac{1}{M_X^2}\cdot
	     \frac{\chi_0}{\pi \langle x\rangle}
	    \exp \left[ - \frac{(1 - \langle x\rangle)^2}
			       {\langle x\rangle} \right]
			       \int^1_{\frac{m_0^2}{M_X^2}}\!\!dx
            \exp \left[ - \frac{(x - \langle x\rangle)^2}
			       {\langle x\rangle} \right] .
			       \label{eq:RESAPP}
\end{equation}
We would like to emphasize two aspects of the approximate formula above: 
\begin{itemize}
\item[$(i)$] it explicitely exhibits the characteristic $M_X^2$ 
dependence of diffractive  collisions, namely it shows the 
$(M_X^2)^{-\alpha}$ behaviour with $\alpha \simeq 1$;
\item[$(ii)$] the exponential and integral factors have a very weak
dependence on $M_X^2$ but they contain a non-trivial energy $ \sqrt{s}$ 
dependence, which is intrinsic to the model and comes ultimately from
phase space limits and cross sections contained in eq.(\ref{eq:OMEGA}). 
\end{itemize}

In Fig. 2 we compare eq.(\ref{eq:RES})
with the recent data from the H1 collaboration 
\cite{HERA}. In all formulas $\sqrt{s}$ will now be replaced by $W$, 
the photon-proton center of mass energy. Figure 2a (2b) presents data 
for $W=187$ GeV ($W=231$GeV). The different curves 
correspond to the choices
I ($m_{0}=0.31$ GeV, $\sigma=2.7$ mb), 
II ($m_{0}=0.35$ GeV, $\sigma=2.7$ mb),
III ($m_{0}=0.31$ GeV, $\sigma=5.4$ mb) and
IV ($m_{0}=0.35$ GeV, $\sigma=5.4$ mb), respectively. In all these 
curves we have used $G_{I\!\!P}(y) = 6 \frac{(1-y)^5}{y}$. 
As expected, the distribution at low $M^2_X$ is very sensitive to 
threshold effects. When we go from the upper to the lower solid 
(dashed) lines we can observe that the increase of the Pomeron-hadron 
cross section changes the distribution in such a way that larger 
masses $M^2_X$ are favoured. 

We would like to stress that in curve II there is no free or 
new parameter. All parameter values are the same as in our 
previous paper devoted to hadronic diffraction. It misses 
only the very small mass region points, where we expect it 
to be below the data, since we do not include resonance
effects. In the large mass region a better agreement with data may be 
achieved with a somewhat larger value of the Pomeron-hadron 
cross section. This region may, however, be influenced by other 
effects, one of which we discuss below.

In Fig. 3 we compare the same data ($W=187$ GeV) 
with our mass spectrum obtained with  $G^h_{I\!\!P}(y)$ (curve I), 
$G_{I\!\!P}(y)$ (curve II) and $G^s_{I\!\!P}(y)$ (curve III). This
comparison suggests that the ``hard'' Pomeron can give a good 
description of data. The same can be said about our ``mixed''
Pomeron, which, in fact seems to be more hard than soft. These three
curves were calculated with exactly the same parameters and 
normalizations, the only difference being the Pomeron profile. 
Apparently the ``soft'' Pomeron (curve III) is ruled out by data.
Curve IV shows, however, that with a different choice of parameters 
$m_{0}=0.50$ GeV and $\sigma=5.4$ mb a good agreement is again
obtained. Considering the large ammount of data already described
previously by the IGM, this choice is extreme. We conclude therefore
that the ``soft'' Pomeron is disfavoured. This same conclusion was
found in refs. \cite{zeus,h1}.

A very interesting question regarding DD processes
is whether or not semihard interactions play a role in 
diffractive physics. In hadronic non-diffractive collisions semihard
scatterings are expected to be visible at c.m.s. energies around
$\sqrt{s}\simeq 500$ GeV. In such scatterings two partons interact
with a momentum transfer of $p_{T} \simeq 2-4$ GeV, forming two 
so-called minijets. Since $\Lambda_{QCD} \ll p_{T} \ll \sqrt{s}$,
minijet cross sections can be calculated with perturbative QCD and 
they are large enough to be relevant for minimum bias
physics. In the IGM, energy deposition is occuring due to gluon-gluon 
collisions in both perturbative (semihard) and non-perturbative 
regimes.  
The gluon-gluon cross section in eq. (\ref{eq:OMEGA}) is computed with
perturbative QCD or with a non-perturbative ansatz according to the
scale ($=xys$).  The relative importance of minijets with respect to
the soft processes was fixed following the experimental estimates of 
the minijet cross section made by the CERN UA1 and UA5 collaborations
in hadronic collisions. We have been assuming so far that the onset
of semihard physics in DD occurs at the same energy as in ordinary
non-diffractive processes. This may not be true, or even if it is true,
there are uncertainties regarding the precise value of the relevant
energy scale. In Fig. 4 we repeat the fit of Fig. 2 
using only curves II, 
which are our ``conservative'' calculation, plotted with solid lines.
 Since the present energies are not yet very large, 
we could just neglect the small minijet component.
The dashed curves in Fig. 4 show the effect of switching off the 
semihard contribution. There is only a small enhancement in the tail of
the spectra. Without minijets the energy deposition in the central
blob of Fig. 1 is decreased and the leading particle, in the
upper leg ($\rho$ meson after interaction), is more energetic. It
contributes more to the diffractive mass $M^2_X$ and makes it larger.
This effect is negligible for very low $M^2_X$ but becomes visible
at large diffractive masses.  Repeating this
comparison (total spectrum versus the spectrum without minijets)
at higher energies we observe that the magnitude of the minijet
contribution is always small and shows always the tendency to
produce slightly faster falling distributions at the end of the 
spectrum. This 
suggests that in DD processes minijets are unimportant even at very
high energies.

\section{Summary and conclusions}

In conclusion, a straightforward (and with no new parameter) extension 
of our model of hadronic diffraction to photon-proton reactions is
able to fit the data for diffractive mass excitation presented 
in ref. \cite{HERA} within small
discrepancies. The agreement may become better with some small
changes motivated by uncertainties in previous fitting procedures.

Our analysis of data suggests that the Pomeron is ``hard'' and
undergoes mostly ``soft'' interactions. This means that this
object is composed by a relatively ``small'' number of gluons 
carrying each, in the average a large fraction of the Pomeron 
momentum, but a small fraction of the total momentum of the  proton 
 and undergoing
mostly soft (with respect to a hard energy scale $\simeq 2-3$ GeV)
collisions with the gluons of the other hadron.

The fact that our model is successful means that energy flow in
many and different high energy reactions can be understood as
an incoherent superposition of parton-parton scatterings constrained
by energy conservation.

\vspace{0.5cm}
\underline{Acknowledgements}: This work has been supported by FAPESP,
CNPQ (Brazil) and KBN (Poland).  We would like to warmly thank 
R. Covolan and E. Ferreira for many fruitful discussions. GW would also like
to thank IFUSP for warm hospitality extended to him during his visit there.


\vspace{1cm}
\noindent
{\bf Figure Captions}\\
\begin{itemize}
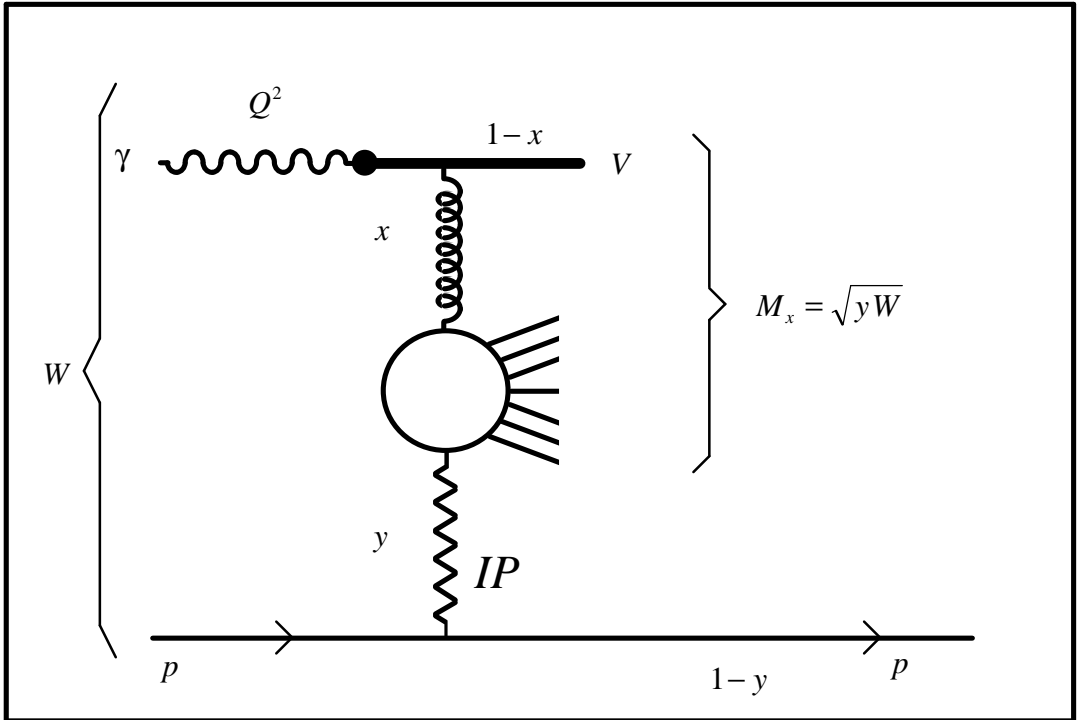
\item[{\bf Fig. 1}] IGM description of a photon-proton scattering
with the formation of a diffractive system of invariant mass $M_X$.

\item[{\bf Fig. 2a}] Diffractive mass spectrum for $\gamma p$
collisions at $W=187$ GeV
 calculted with the IGM (eq.(\ref{eq:RES})) and compared
with H1 data \cite{HERA}. 
The different curves correspond to the choices:
I ($m_{0}=0.31$ GeV, $\sigma=2.7$ mb), 
II ($m_{0}=0.35$ GeV, $\sigma=2.7$ mb),
III ($m_{0}=0.31$ GeV, $\sigma=5.4$ mb) and
IV ($m_{0}=0.35$ GeV, $\sigma=5.4$ mb), respectively. 

\item[{\bf Fig. 2b}] The same as Fig. 2a for $W=231$ GeV.

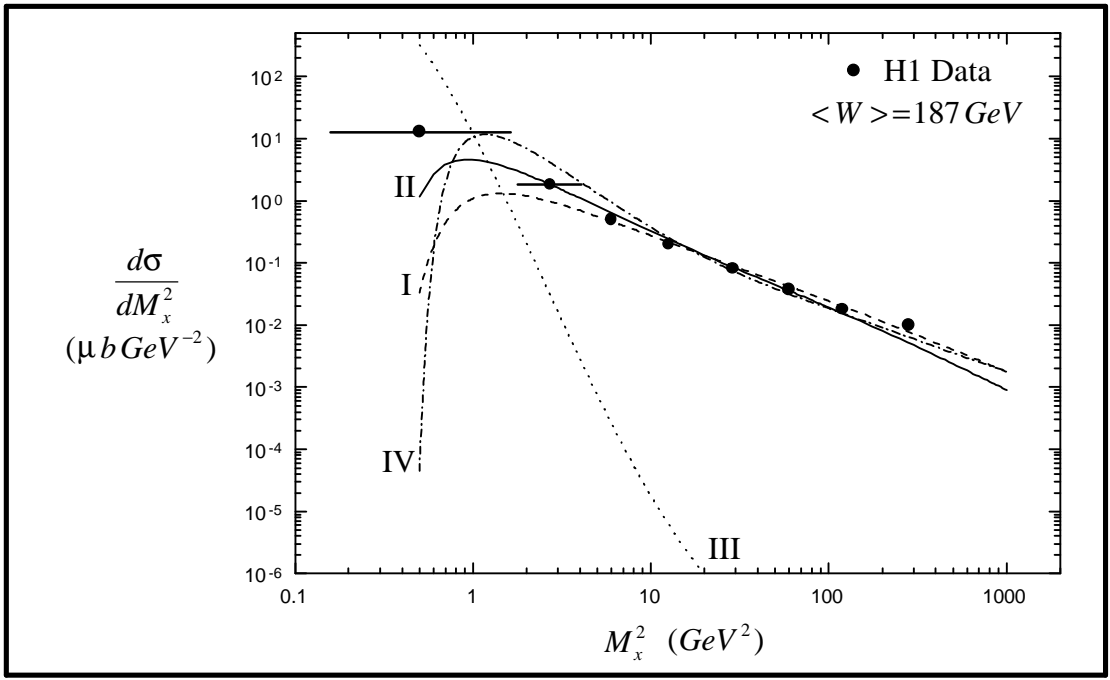
\item[{\bf Fig. 3}] Data from ref. \cite{HERA} compared with eq. 
(\ref{eq:RES}). The solid line (curve II) corresponds to the
choice $m_{0}=0.35$ GeV, $\sigma=2.7$ mb and $G_{I\!\!P}(y)$. 
Curves I (dashed) and III (dotted) are obtained replacing 
$G_{I\!\!P}(y)$ by $G^h_{I\!\!P}(y)$ and $G^s_{I\!\!P}(y)$ 
respectively. Curve IV is obtained with $G^s_{I\!\!P}(y)$
and $m_{0}=0.50$ GeV and $\sigma=5.4$ mb. 

\item[{\bf Fig. 4a}] Diffractive mass spectrum for $\gamma p$
collisions at $W=187$ GeV, curve II of Fig. 2a, shown with a
solid line and compared with the same spectrum without the
minijet contribution (dashed line).

\item[{\bf Fig. 4b}] The same as Fig. 3a for $W=231$ GeV.

\end{itemize}


\begin{thebibliography}{99}
\bibitem{FOOT} It can be also deflected with some invariant momentum
               transfer $t$. However, because the IGM does not 
               include so far any transverse momentum transfers,
               the results presented here must be regarded as
               appropriately averaged/integrated over $t$.
\end{thebibliography}
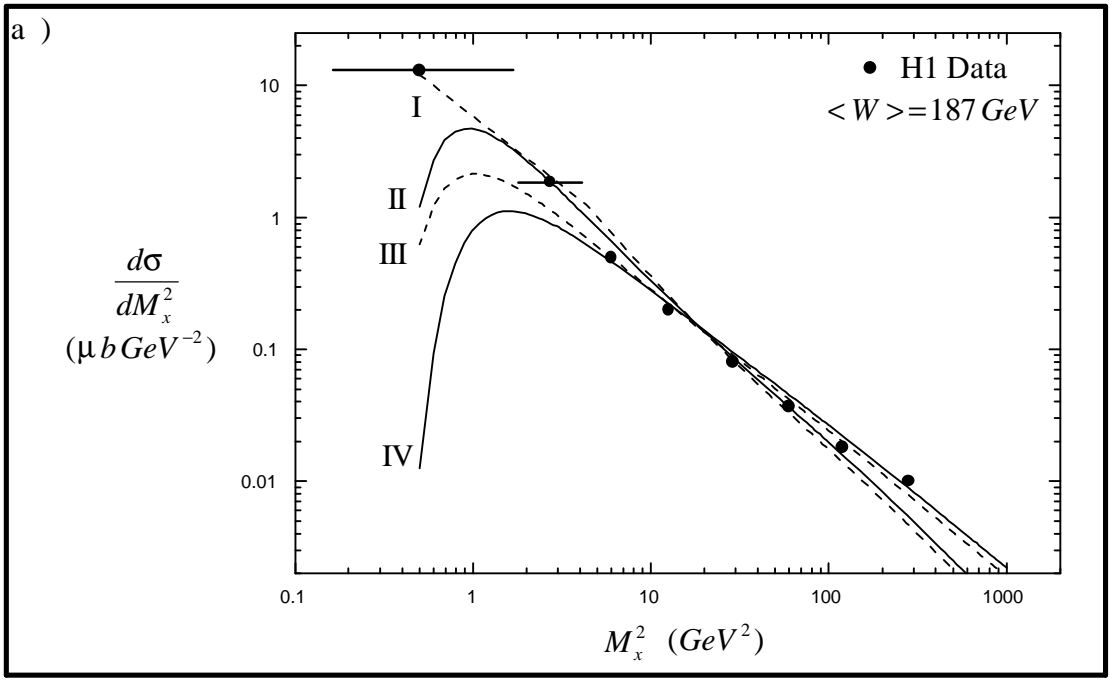
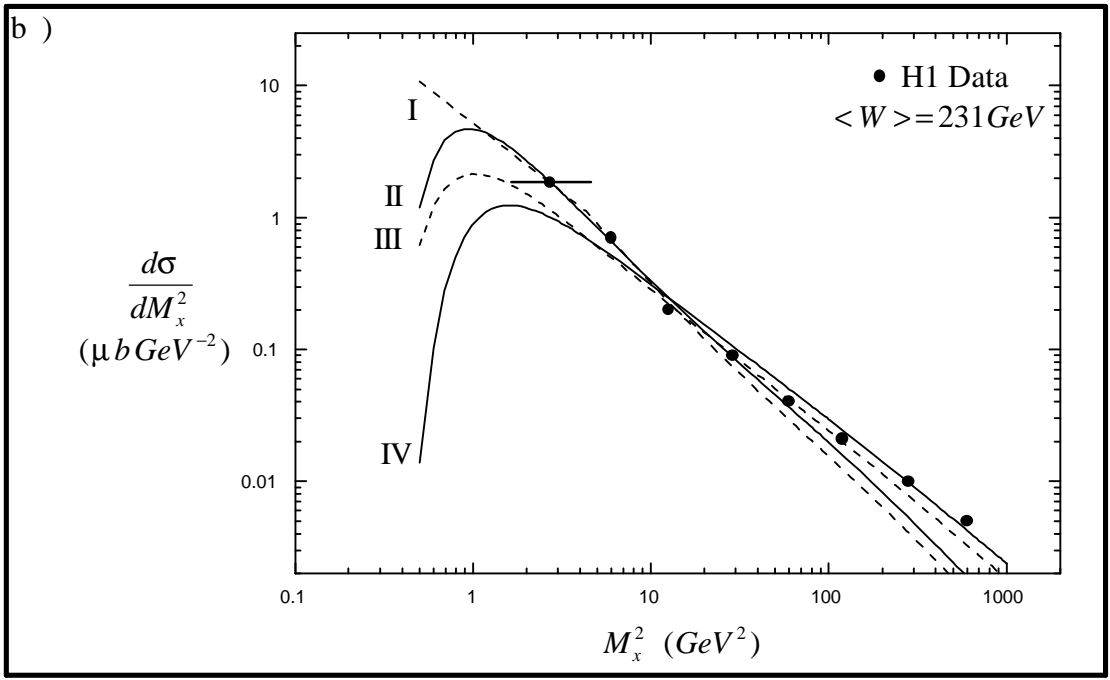
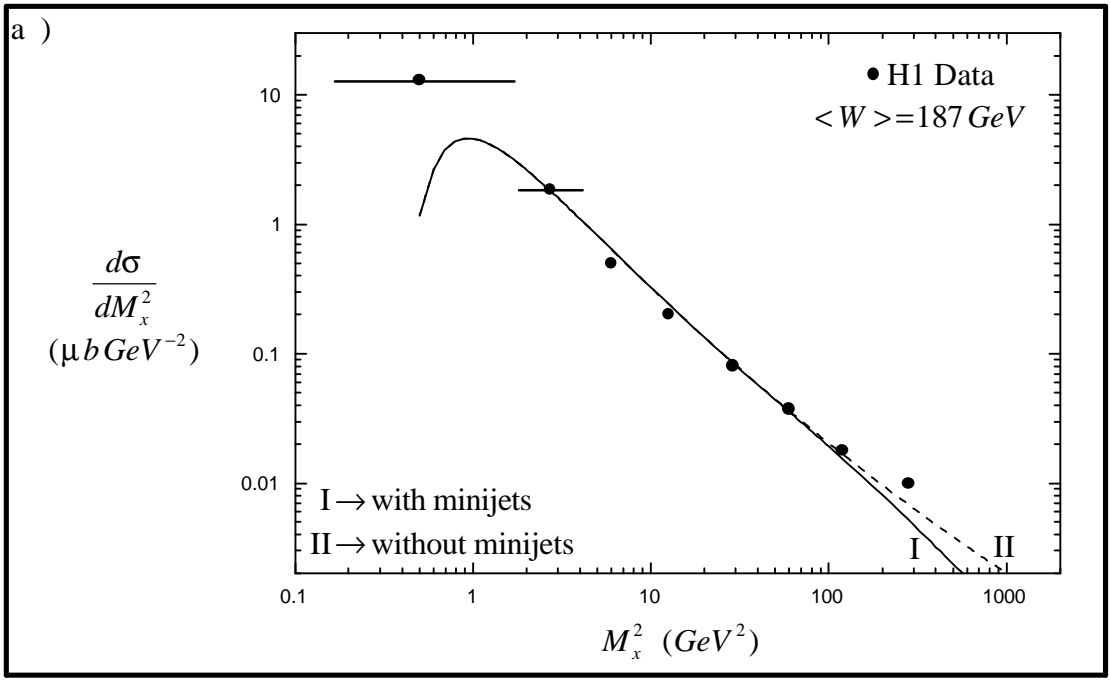
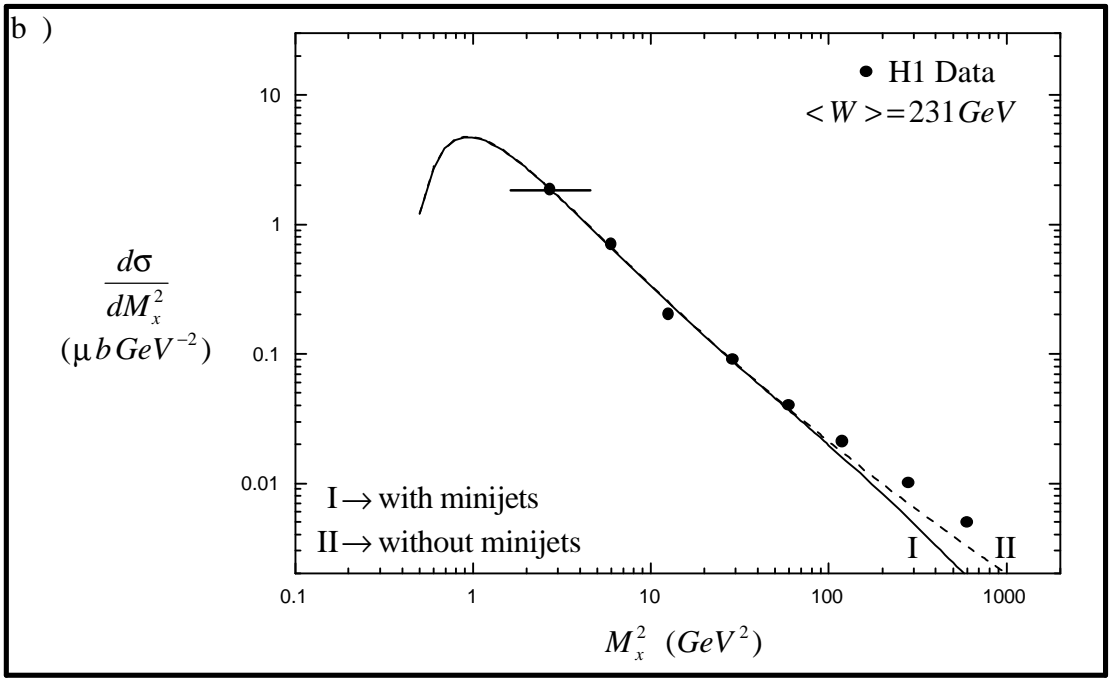
\end{document}